\documentclass[12pt]{article}
\usepackage{amsmath}
\usepackage{amsfonts}
\usepackage{amssymb}
\newcommand{\be}{\begin{equation}}
\newcommand{\ee}{\end{equation}}

\newcommand{\no}{\nonumber\\}
\newcommand{\ba}{\begin{eqnarray}}
\newcommand{\ea}{\end{eqnarray}}
\newcommand{\ci}[1]{\cite{#1}}
\newcommand{\bi}[1]{\bibitem{#1}}
\newcommand{\la}[1]{\label{#1}}
\def\gl#1{(\ref{#1})}

\date{}
\begin{document}
\title{Localization of scalar fields on self-gravitating thick branes}
\author{
Alexander A. Andrianov$^{ab}$, Vladimir A. Andrianov$^{a}$, Oleg O. Novikov$^{a}$
\\
$^{a}$ V.A. Fock Department of Theoretical Physics,\\\phantom{$^{a}$} Sankt-Petersburg State University,\\
\phantom{$^{a}$}
ul. Ulianovskaya, 198504 St. Petersburg, Russia\\
$^{b}$ Institut de Ci\`encies del Cosmos, Universitat de Barcelona,\\ \phantom{$^{a}$}
Marti i Franques, 1, 08028
Barcelona, Spain
\\
{\small E-mail: {\tt andrianov@icc.ub.edu, v.andriano@rambler.ru, oonovikov@gmail.com}}}
\maketitle

\begin{abstract} The model of a domain wall ("thick" brane) in
noncompact five-dimensional space-time  is considered with geometries of
$AdS_5$ type generated by self-interacting scalar matter .  The scalar matter is composed of two fields with $O(2)$ symmetric self interaction. One of them is mixed with gravity scalar modes and plays role of the brane formation mode (due to a kink background) and another one is of a Higgs-field type. The interplay between  soft breaking of $O(2)$ symmetry and gravity influence is thoroughly investigated around the critical point of spontaneous $\tau$ symmetry breaking when the v.e.v. of the Higgs-type scalar field occurs. The
possibility of (quasi)localization of scalar modes on such thick branes is examined.\end{abstract}
\section{Introduction}
Recent years, the models based on the hypothesis that our universe
is a four-dimensional space-time hypersurface (3-brane) embedded in
a fundamental multi-dimensional space \ci{rushap},\ci{otherbr} have become quite popular,
see, for example, the reviews  \ci{RuBar}-\ci{loc6} and the references
therein. The number of extra dimensions, their characteristic size
and the number of physical fields, which are spread out the bulk
space,  may be different in  various approaches. At the same time it
is assumed that the additional space size is large enough, and
additional dimensions can be, in principle, detected in terrestrial
experiments planned in the near future and/or in astrophysical
observations. Four dimensions of our world can be ensured, in
particular, by the localization mechanism of matter fields on
three-dimensional hypersurfaces in multidimensional space, i.e.
3-branes. Different scenarios of domain walls description and their
applications to elementary particle physics and cosmology can be
found in a number of reviews \ci{rev1} - \ci{loc6}. The influence
of  gravity is especially interesting, which plays an important role
in a (quasi)localization of matter fields on the brane \ci{rev12} -
\ci{singl1}. The question arises, under what
circumstances the (quasi)localization of matter fields with spin zero on a
brane is still possible  when the  minimal interaction with gravity
is present? This work is devoted partially to answer this question.

In this paper we consider a model of the domain wall formation with
finite thickness ("thick" branes ) by self-interacting scalar fields and
gravity in five-dimensional noncompact space-time \ci{aags2} with  anti-de Sitter geometries
on both sides of the brane.
The formation of  "thick" brane with the localization of light
particles on it was obtained earlier in \ci{aags1} with the help of a
background scalar and  the gravitational fields, when their
vacuum configurations have nontrivial topology. Appearance of scalar
states with (almost) zero mass on a brane has happened to be
possible. However, as it was previously shown \ci{rev19}, the
existence of the centrifugal potential in the second variation of scalar-field action may lead to  absence of
localized modes on a brane.

In the present work the scalar matter is composed of two fields with $O(2)$ symmetric self interaction. One of them ("branon" \cite{branon}) is mixed with gravity scalar modes and plays role of the brane formation mode (due to a kink background) and another one is a fermion mass generating (FMG) field (replacing a Higgs field). The soft breaking of $O(2)$ symmetry by tachyon mass terms for both fields is introduced which eventually generates spontaneous breaking of translational symmetry due to formation of kink-type field v.e.v. Furthermore for special values of tachyon mass terms the critical point of spontaneous $\tau$ symmetry breaking exists when the v.e.v. of the FMG scalar field occurs. In the first phase the only nontrivial v.e.v. is given by a kink configuration. But the branon fluctuations around kink in the presence of gravity are suppressed by the universal repulsive centrifugal potential which survives in the zero gravity limit \ci{rev19}. Thus gravity induces a discontinuity in the branon field spectrum. However the FMG field in this phase decouples from branons, is massive and exhibits a more regular weak gravity behavior.
In the second phase the Higgs-type field obtains a localized v.e.v. to be used for generation of fermion masses \ci{aags1}. Both fields, branons and FMG scalars, are mixed and the scalar mass spectrum and eigenstates must be found by functional matrix diagonalization.

The work starts (Section 2) with brief motivation of necessity for two scalar fields to provide fermion localization on domain wall \ci{loc1} -\ci{loc5} and to supply localized Dirac fermions with masses. In Section 3 the model of two scalar fields  with their minimal
coupling to gravity is formulated for arbitrary potential and the equations of motion are derived.  In the subsection 3.2 the scalar potential is restricted  with a quartic $O(2)$ symmetric potential
and soft breaking of  $O(2)$ symmetry quadratic in fields ( as it could arise from the fermion induced effective action \ci{aags2}). For this Lagrangian the gaussian normal coordinates are introduced and the appropriate equations of motion are obtained. The existence of two phases  which differ in presence or absence of v.e.v. for the FMG field is revealed and the solutions for classical background of both scalar fields are found  in the leading approximation of the gravity coupling expansion. In the  Subsection 3.3 the next-to-leading approximation is performed.

 In the
section 4 the full action is derived up to quadratic order in
fluctuations in a vicinity of a background metric. It is
dedicated to the separation of equations in respect to different
degrees of freedom. At the end of this section the action of scalar fields for the brane and
gravity  is obtained in gauge invariant variables. In
sect. 5  the mass
spectrum for the "thick" brane in the theory with a quartic $O(2)$ symmetric potential
and soft breaking of  $O(2)$ symmetry quadratic in fields  is investigated around the critical point in the weak gravity expansion. In conclusion, we discuss results and
prospects of the proposed model.
\section{Motivation of the two scalar field model}
Let start with elucidating how to trap fermion matter on a domain wall -- "thick brane".
The latter one
emerges in the model of
five-dimensional fermion bi-spinors $\psi(X)$
coupled to a scalar field $\Phi(X)$.
The extra-dimension coordinate is assumed to be space-like,
$$
(X_\alpha) = (x_\mu, z)\ , \quad (x_\mu) = (x_0, x_1, x_2, x_3)\ , \quad
(\eta_{\alpha\alpha}) = (+,-,-,-,-)
$$
and the subspace of $x_\mu$ corresponds to the
four-dimensional Minkowski
space. The extra-dimension size is assumed to be infinite (or large enough).
The fermion wave function  then obeys by the Dirac equation
\be
[\,i\gamma_\alpha \partial^\alpha - \Phi(X)\,]\psi(X) = 0\ , \quad
\gamma_\alpha = (\gamma_\mu, -i\gamma_5)\ ,\quad \{\gamma_\alpha,
\gamma_\beta\}
= 2\eta_{\alpha\beta}\ , \la{5dir}
\ee
with $\gamma_\alpha$ being a set of four-dimensional Dirac matrices in the
chiral representation.

The trapping of
light fermion states on a four-dimensional hyper-plane -- the domain wall -- the "thick brane"
is provided by localization mechanism in the
fifth dimension at $z = z_0$. It is facilitated
by a certain $z$-dependent background configuration of
the scalar field  $\langle\Phi(X)\rangle_0 = \varphi (z)$,
which provides the appearance of zero-modes in the four-dimensional fermion spectrum.
For the four-dimensional space-time interpretation,
Eq.\,\gl{5dir} can be decomposed into the infinite set of fermions
with different masses calculable from the following squared Dirac equation,
\ba
&&[\,i\gamma_\alpha \partial^\alpha + \varphi(z)\,][\,i\gamma_\alpha
\partial^\alpha - \varphi(z)\,]\psi(X)
\equiv (- \partial_\mu \partial^\mu - \widehat m^2_z) \psi(X)\ ;\no
&&\widehat m^2_z = - \partial_z^2 + \varphi^2 (z) - \gamma_5 \varphi'(z) =
\widehat m^2_{+} P_L + \widehat m^2_{-} P_R\ , \la{f-s}
\ea
where  $P_{L,R} = \frac12 (1 \pm \gamma_5)$ are projectors on
the left- and right-handed states.
Thus the mass squared operator $\widehat m^2_z$ consists of two chiral partners
\ba
\widehat m_\pm^2 &=&  - \partial_z^2 + \varphi^2 (z) \mp \varphi'(z)
=  [\,-\partial_z \pm \varphi(z)\,][\,\partial_z \pm \varphi(z)\,]\ ;
\la{fact}\\
\widehat m_{+}^2\,q^+ &=& q^+\,\widehat m_{-}^2,\quad
\widehat m_{-}^2\,q^- = q^-\,\widehat m_{+}^2\ ,\quad q^\pm \equiv
\mp \partial_z + \varphi(z)\ . \la{susy}
\ea
Due to such a supersymmetry \ci{susy1}-\ci{susy3}, for non-vanishing masses, the
left- and right-handed spinors in \eqref{susy} form the bi-spinor describing
a dim-4 massive Dirac particle which is, in general, not  localized at any
point of the extra-dimension for asymptotically constant
field configurations $\varphi(z)$. Such a spectral equivalence may be broken by
a normalizable zero mode of
one of the mass operators $\widehat m_\pm^2$. This mode is read out of  Eqs.\ \gl{fact} and \gl{susy}
\be
q^-\psi^{+}_0(x,z) = 0\ , \quad \psi^{+}_0(x,z) =
\psi_L(x) \, \exp\left\{-\int^z_{z_0} dw\varphi(w)\right\}\ ,
\ee
where $\psi_L(x) = P_L \psi (x)$ is a free-particle
Weyl spinor in the four-dimensional Minkowski space.
Evidently, if a scalar field configuration has
the appropriate asymptotic behavior,
$$
\varphi(z)\stackrel{z \rightarrow \pm\infty}{\sim}
\pm C_\pm |z|^{\nu_\pm}\ ,\quad \mbox{\rm Re} \nu_\pm > -1\ ,
\quad  C_\pm > 0\ ,
$$
then the wave function $\psi^{+}_0(x,z)$ is normalizable
on the $z$ axis and the corresponding
left-handed fermion is a massless Weyl particle localized in the vicinity
of a four-dimensional domain wall.  If
$\varphi(z)$ is asymptotically constant, with $ C_\pm > 0$ and $\nu_\pm = 0$ then
there is a gap for the massive Dirac states.

In this paper we restrict ourselves with generating parity symmetric branes by field configurations of definite parity.  The example of  a parity odd topological configuration  is realized by
a kink-like scalar field background (of possibly dynamical origin, see below)
\be
\varphi^{+} =
M\, \mbox{\rm tanh}(Mz)\ . \la{soli}
\ee
The two mass operators have the following potentials
\be
\widehat m^2_{+} =- \partial_z^2 + M^2
\left[\,1-2{\rm sech}^2(Mz)\,\right];\quad
\widehat m^2_{-} =- \partial_z^2 + M^2, \la{chpot}
\ee
and the left-handed normalized zero-mode is localized around $z=0$,
\be
\psi^{+}_0(x,z) =
\psi_L(x)\,\psi_0 (z)\ ,\qquad \psi_0 (z) \equiv
\sqrt{M/2}\ {\rm sech}(Mz)\ . \la{locmod}
\ee
Evidently
the threshold for the continuum is at $ M^2$
and the heavy Dirac particles may have
any masses $m > M$. The corresponding
wave functions are  spread out in the fifth dimension.

But the fermions of the Standard Model are mainly massive and composed of both left- and right-handed spinors. Therefore, for  light fermions
on a brane one needs
at least two five-dimensional fermions $\psi_1(X), \psi_2(X)$
in order to generate left- and right-handed
parts of a four-dimensional Dirac bi-spinor as zero modes.
The required zero modes with different chiralities for
$\langle\Phi(X)\rangle_0 = \varphi^+(z)$ arise when the two fermions couple
to the  scalar field $\Phi(X)$ with opposite charges,,
\be
[\,i\not\!\partial - \tau_3\Phi(X)\,]\Psi(X) = 0\ ,\quad
\not\!\partial \equiv \widehat\gamma_\alpha \partial^\alpha\ ,\quad
\Psi(X) =\left\lgroup\begin{array}{c}\psi_1(X)\\
\psi_2(X)\end{array}\right\rgroup\ ,
\la{2fer}
\ee
where $\widehat\gamma_\alpha \equiv \gamma_\alpha\otimes {\bf 1}_2$ are
Dirac matrices and
$\tau_a \equiv {\bf 1}_4 \otimes \sigma_a,\ a=1,2,3 $
are the generalizations of
the Pauli matrices $\sigma_a$ acting on the bi-spinor components $\psi_i(X)$.

In this way one obtains a massless Dirac particle on the brane
and the next task
is to supply it with a light mass. As the mass operator
mixes left- and right-handed
components of the four-dimensional fermion it is embedded in the Dirac
operator \gl{2fer} with the mixing matrix $\tau_1 m_f$ of the fields
$\psi_1(X)$ and $\psi_2(X)$.
If realizing the Standard Model mechanism
of fermion mass generation by means of dedicated scalars,
one has to introduce the
second scalar field $H(x)$, replacing the bare mass
$\tau_1 m_f \longrightarrow \tau_1 H(x)$ in the Lagrangian density \cite{aags1},
\ba
{\cal L}^{(5)} (\overline{\Psi},\Psi,\Phi, H) =
\overline{\Psi} ( i\!\not\!\partial
- \tau_3 \Phi -\tau_1 H) \Psi . \la{aux}
\ea Both scalar fields may
be dynamical and their self-interaction should justify the spontaneous
symmetry breaking by certain classical configurations
trapping light massive fermions on the domain wall. If the lagrangian of scalar fields is symmetric under reflections $\Phi,  -\,\Phi$ and $H \longrightarrow -\,H$ then the invariance may hold under discrete $\tau$-symmetry
transformations,
\ba
&&\Psi \longrightarrow \tau_1 \Psi\ ;\quad
\Phi \longrightarrow -\,\Phi\ ;\\
&&\Psi \longrightarrow \tau_2 \Psi\ ;\quad
\Phi, H \longrightarrow -\,\Phi, -\,H\ ;\\
&&\Psi \longrightarrow \tau_3 \Psi\ ;\quad
H \longrightarrow -\,H\ ,
\la{tausim}
\ea
the $\tau_2$ symmetry in fact can be extended to the continuous $U_\tau(1)$ symmetry under rotations,
\be
\Psi \longrightarrow \exp\{i \alpha \tau_2/2\} \Psi ;\quad \Phi \longrightarrow \cos\alpha \Phi +\sin\alpha H,\quad H \longrightarrow - \sin\alpha \Phi +\cos\alpha H , \label{utau}
\ee
which could be a high-energy symmetry if the scalar field lagrangian adopts it for large values of fields. But  at full these symmetries
do not allow the fermions to acquire
a mass unless translational invariance is spontaneously broken in the scalar sector.

There may be several patterns of the partial $\tau$ symmetry breaking by scalar field backgrounds. The first one is generated by a $z$-inhomogeneous v.e.v. of only one of the fields, say, the field $\Phi(z)$ with $H (z) = 0$ . Then the $\tau_3$ symmetry certainly survives but the $\tau_{1,2}$ symmetries are broken. Still  if the function $\Phi(z)$ is odd against reflection in $z$ the latter symmetries can be restored being supplemented by reflection $z \longrightarrow - z$ ($\tau_3 P$(arity) symmetries). The second pattern is supported by $z$-inhomogeneous v.e.v.'s of both scalar fields provided that $\Phi(z) \not\sim H(z)$. Then, in general, none of the $\tau$ symmetries holds. But  if  $\Phi(z)$ and $H (z)$ are odd and even functions respectively, the $\tau_3 P$ symmetry may again survive.

Thus one may anticipate a phase transition between the phases with different symmetry patterns which is presumably of the second order if the v.e.v. $H (z)$ is continuous in coupling constants of the model. This realization is welcome to implement  light fermion masses near a phase transition which are governed by a small deviation in parameters of the scalar field potential around a scaling point much less than the localization scale $M$.

Further on we assume that the dynamics of fermions and scalar fields is $\tau$- and $U_\tau(1)$-symmetric \eqref{tausim}, \eqref{utau} at high energies whereas at low energies $U_\tau(1)$-symmetry is broken softly and $\tau$-symmetry is violated spontaneously. Accordingly the scalar field potential contains even powers of fields $\Phi(z)$ and $H (z)$ and its profile induces the required spontaneous symmetry breaking. A concrete model for two phases with broken translational invariance is presented in the next section.

\section{Formulation of the model in bosonic sector}
\subsection{General two-boson potentials: conformal coordinates}
Eventually we want to examine  the properties of scalar matter generating  gravity. Therefore
let us supply the five-dimensional space with gravity providing it with a pseudo Riemann metric tensor
 $g_{AB}$. This tensor in flat space and for the rectangular coordinate system is reduced to $\eta^{AB}$. We define the dynamics of two real scalar fields $ \Phi(X)$ and $ H(X)$ with a minimal interaction to gravity by the following action functional,
    \be
S[g, \Phi,  H ] = \int {d^5 X}\sqrt {\left| g \right|} {\cal L}(g, \Phi,  H ), \label{1} \ee
    \be
{\cal L} =  \left\{ { - \frac12 M_\ast ^3 R  + \frac12 (\partial _A  \Phi \partial ^A  \Phi +\partial _A  H \partial ^A  H)  - V\left(  \Phi,  H  \right)} \right\}, \ee where $R$ stands for a scalar curvature, $ \left| g \right| $ is the determinant of the metric tensor, and $ M_\ast $ denotes a five-dimensional gravitational Planck scale.

The equations of motion are
    \ba&&
R_{AB}  - \frac{1}{2}g_{AB} R = \frac{1}{{M_\ast^3 }}T_{AB} ,\no    &&  D^2  \Phi  =  - \frac{{\partial V}}{{\partial  \Phi }} ,\quad  D^2  H  =  - \frac{{\partial V}}{{\partial  H }},
 \ea where $D^2$ is a covariant D'Alambertian, and the energy-momentum tensor reads,
    \be
T_{AB}  = \partial _A  \Phi \partial _B  \Phi + \partial _A  H \partial _B  H - g_{AB} \left(\frac12 {\partial _C  \Phi \partial ^C  \Phi + \partial _C  H \partial ^C  H  - V\left(  \Phi,  H  \right)} \right). \ee
In order to build a thick $3 + 1$-dimensional brane we  study such classical vacuum configurations which do not violate spontaneously 4-dimensional Poincare invariance. In this Section  the metric is represented in the conformally flat form, $g_{AB} = A^2 \left(z \right) \eta_{AB} $. This kind of metric suits well for interpretation of scalar fluctuation spectrum and their resonance effects (i.e. scattering states).

For this metric the equations of motion read,
 \ba &&\left(\frac{A'}{A^2}\right)'= - \frac{ \Phi'^2 +  H'^2}{3 M^3_\ast A},\quad -2A^5 V( \Phi,  H) =  3 M^3_\ast \Bigl( A^2A'' + 2 A (A')^2\Bigr), \label{eoms0}\\ &&\left(A^3 \Phi'\right)'= A^5\frac{\partial V}{\partial \Phi} ,
\quad \left(A^3  H'\right)'= A^5\frac{\partial V}{\partial  H}. \label{eoms}\ea
One can prove \ci{aags1}, that only three of these equations are independent.

Following the arguments of the previous section we assume that the potential is analytic in scalar fields, exhibits the discrete symmetry under reflections $ \Phi \longrightarrow  -\, \Phi$ and $ H \longrightarrow -\, H$ and has a set of minima  for nonvanishing v.e.v. of scalar fields.
Correspondingly there exist constant background solutions $\{ \Phi_{min},\  H_{min}\}$ which are compatible  with the Einstein equations provided that $\langle V ( \Phi,  H)\rangle = V(\{ \Phi_{min},\  H_{min}\}) \equiv  \lambda_{cosm} M^3_\ast < 0$, i.e. for positive cosmological constant $\lambda_{cosm}$. In this case the warped geometry will be of Anti-de-Sitter type, $1/A \sim \pm k z$ with AdS curvature $k = \sqrt{-\lambda_{cosm}/6}$ as in the Randall-Sundrum model II \cite{RSII} .

\subsection{Minimal realization in $\phi^4$ theory: gaussian normal coordinates }
In this Subsection we study the formation of a brane in the theory with a minimal stable potential admitting kink solutions. It possesses a quartic scalar self-interaction
and  wrong-sign mass terms for both scalar fields  This potential is designed with  $U_\tau(1)$-symmetry of dim-4 vertices but with different quadratic couplings. The conveniently  normalized effective action has the form, \ba &&S_{eff}(\tilde\Phi ,g) = \frac12 M_\ast^3\int {d^5 X} \sqrt {|g|} \Bigl\{  -R +2
\lambda_{cosm}  +\frac{3\kappa}{M^2}\Big(\partial _A \tilde\Phi \partial ^A \tilde\Phi +\partial _A \tilde H \partial ^A \tilde H \no &&\phantom{S_{eff}(\Phi ,g) = \frac12 M_\ast^3\int {d^5 X} \sqrt {|g|}} + 2M^2 \tilde\Phi^2 +2\Delta_H \tilde H^2- (\tilde\Phi^2 + \tilde H^2)^2 -\tilde V_0\Big)  \Bigr\}, \label{36} \ea
where the normalization of the kinetic term of scalar fields $\kappa $ is chosen differently from \eqref{1} in order to simplify the Eqs. of motion (see below)\footnote{ It could be inherited from the low-energy effective action of
composite scalar fields induced by the one-loop dynamics  of five-dimensional pre-fermions \cite{aags2}.}. They are connected as follows,
\be \Big[ \Phi,  H\Big] = \left(\frac{3\kappa M^3_\ast}{M^2}\right)^{1/2} \Big[\tilde\Phi, \tilde H\Big]. \label{variab}\ee
For relating it to the weak gravity limit we guess that $ \kappa \sim M^3/M_\ast^3 $ is a small parameter, which
characterizes the interaction of gravity and matter fields. Let us take $M^2 > \Delta_H $ then the true minima are achieved at $\tilde\Phi_{min} = \pm M,\ \tilde H_{min} = 0 $ and a constant shift of the potential energy must be set $V_0 = M^4$ in order to determine properly the cosmological constant $\lambda_{cosm}$.

Now we change the coordinate frame to the warped metric in gaussian normal coordinates, \be ds^2  =
\exp \left( { - 2\rho \left( y \right)} \right)dx_\mu  dx^\mu   -
dy^2 ,\quad y = \int^z_0 dz' A(z').\label{met}\ee
 This choice happens to be more tractable for analytic calculations than the conformal one used for \eqref{eoms}. With the definition \eqref{met} the function $y(z)$ is monotonous and  $z \rightarrow - z \Longrightarrow y \rightarrow - y $.

The Eqs. of motion \eqref{eoms} for this metric take the form, \ba &&\tilde\Phi'' =  - 2M^2\tilde\Phi  + 4\rho'\tilde\Phi' + 2\tilde\Phi(\tilde\Phi^2 + \tilde H^2) ,\label{38}\\
&& \tilde H'' =  - 2 \Delta_H \tilde H  + 4\rho' \tilde H
' + 2 \tilde H(\tilde\Phi^2 + \tilde H^2) ,\label{39}\\ &&\rho''   = \frac{{\kappa }}{{M^2
}}\ (\tilde\Phi'^2 + \tilde H'^2), \label{40A}\\   &&\lambda_{cosm} = - 6\rho '^2  + \frac{3{\kappa
}}{{2M^2 }}\left\{ (\tilde\Phi)'^2  +(\tilde H')^2 + 2M^2\tilde\Phi ^2 + 2\Delta_H \tilde H^2 - (\tilde\Phi^2 + \tilde H^2)^2 - M^4 \right\}
.\label{401}
\ea
When compared to Eqs. \eqref{eoms} one finds that in the gaussian coordinates the equations \eqref{38}, \eqref{39}, \eqref{40A} are algebraically simpler being linear in the metric factor $\rho(y)$. It allows to calculate few first orders in gravitational perturbation theory analytically.

As expected for constant background solutions $\tilde\Phi_{min} = \pm M,\ \tilde H_{min} = 0 $ the cosmological constant $\lambda_{cosm}$ completely determines the metric factor $\rho' = \sqrt{-\lambda_{cosm}/6}$.
In general, for any classical solution, the right-hand side of (\ref{401}) is an integration constant  that can be proven by
differentiating this equation. Thus $\lambda_{cosm}$ is indeed a true constant at the classical level.

The above equations contain terms which have different orders in small
parameter  $ \kappa $, and accordingly  they can be solved by
perturbation theory assuming that,
    \[
\frac{{\left| {\rho '(y)} \right|}}{M} = O(\kappa ) = \frac{{\left|
{\rho ''(y)} \right|}}{{M^2 }}.
\]
Then in the leading order in $ \kappa $ the equations for the fields $
\tilde\Phi \left(y \right) , \tilde H  \left(y \right) $ do not contain the metric factor,
 and the metric is completely governed  by matter order by order in $\kappa$.

Depending on the relation between quadratic couplings $M^2$ and $\Delta_H$ there are the two types of $z$-inhomogeneous  solutions of the
equations (\ref{401}) which have  the form of a two-component kink \cite{aags1}. {\it For gravity switched off} the first one holds for $\Delta_H \leq M^2/2$,
    \be
\tilde\Phi \rightarrow \Phi_0  = \pm M\tanh \left( {My} \right) + O\left( {\kappa } \right),\quad \tilde H(y) = 0 , \label{kink0}\ee
and therefore
the conformal factor to the leading order in $\kappa$ reads,
    \be
\rho_1 \left( y \right) = \frac{2{\kappa }}{3}\left\{ {\ln \cosh
\left( {M y} \right)+\frac{1}{4}  \tanh^2( M y)}\right\}+ O\left(
{\kappa ^2 } \right) , \label{rho0}\ee
which is chosen to be an even function of $y$ in order to preserve the remaining $\tau$ symmetry.

The second one arises only when $M^2/2 \leq \Delta_H \leq M^2$, i.e. $2\Delta_H = M^2 +\mu^2, \ \mu^2 < M^2$,
\be
\Phi_0(y)  = \pm M\tanh \left( {\beta M y} \right), \quad H_0 (y) = \pm \frac{\mu}{\cosh \left( {\beta M y} \right)} ,\quad \beta = \sqrt{1 - \frac{\mu^2}{M^2}}, \label{zeroap} \ee
wherefrom one can find the conformal factor to the leading order in $\kappa$ in the following form,
    \be
\rho_1 \left( y \right) = \frac{{\kappa }}{3}\left\{ \left(3 -  \beta^2\right) {\ln \cosh
\left( {\beta M y} \right)+\frac{1}{2} \beta^2 \tanh^2(\beta M y)}\right\}+ O\left(
{\kappa ^2 } \right) , \label{rho1}\ee
as well symmetric against $y \rightarrow - y$.
One can see that the asymptotic AdS curvature $k$ (defined in the limit $y \gg 1/M$ when $\rho (y) \sim k y$) is somewhat different in the $\tau$ symmetry unbroken and broken phases,
\be k_{unbroken} = \frac23 \kappa M \quad\mbox{vs.}\quad k_{broken} = \frac23 \kappa M \Big(1 + \frac{\mu^2}{2M^2}\Big)\sqrt{1 - \frac{\mu^2}{M^2}} < k_{unbroken}. \label{asymp}
\ee

As the scalar potential is invariant under reflections $\tilde\Phi(y) \longrightarrow - \tilde\Phi(y)$ and $\tilde H (y)\longrightarrow - \tilde H(y)$ one finds replicas of the kink-type solutions which can be uniquely selected out from coupling to fermions if to specify their chirality ($+M$ for left-handed ones) and the sign of induced masses ($+\mu$ for positive masses).
Let's choose the positive signs further on.

Evidently the second solution generates the fermion mass in \eqref{aux} whereas the first kink leaves fermions massless. The solution breaks $\tau$ symmetry and is of main interest for our model building. Thus there are two phases with different scalar backgrounds and it can be shown (see below) that if  $\Delta_H < M^2/2$ the first kink provides a local minimum but for some $M^2/2 < \Delta_H < M^2$ it gives a saddle point whereas the second kink with $\tilde H\not= 0$ guarantees a local stability.
\subsection{Relationship to conformal coordinate metric\label{confgauss}}
To the leading order in $\kappa$ one can derive a simple relation between conformal factor $A(z)$ and $\rho_1(y)$. Namely, with a certain  ansatz for $A(z)$ the first equation for the metric factor in \eqref{eoms0} taken in the variables \eqref{variab} is linearized,\be
A(z) = \frac{1}{1 +f(z)},\,\, f(0) = 0;\qquad f'' = \kappa \frac{(\tilde\Phi')^2 + (\tilde H')^2}{M^2}\Big(1+ f\Big) . \label{conffac} \ee
Then the expansion in powers of gravitational coupling constant $\kappa$ is given by $f =  \sum_{n=1}^\infty \kappa^n f_n$ and the leading order $f_1$ obviously coincides in functional dependence with \eqref{40A} for the $\tau$-symmetry unbroken phase or \eqref{rho1} for the broken phase,
\be
\kappa f_1(z) = \rho_1(y\rightarrow z) = \frac{{\kappa }}{3}\left\{ \left(3 -  \beta^2\right) {\ln \cosh
\left( {\beta M z} \right)+\frac{1}{2} \beta^2 \tanh^2(\beta M z)}\right\}+ O\left(
{\kappa ^2 } \right) .
\ee
However the perturbative expansion in $\kappa$ is not valid for any
 $z$ . Indeed for  \mbox{$\beta Mz \gg 1/\kappa \gg 1$} the asymptotic $f(z)$ is linearly growing and the second term in the right-hand side of Eq. \eqref{conffac} dominates over the first one which generated the perturbation series.
 In spite of that in conformal  reference frame  the coordinate asymptotic at $Mz \gg 1$  is given by the leading order in $\kappa$ as $f(z) \rightarrow k z$ the  next orders in $\kappa$ have a more complicated nonanalytic structure. Thereby  the perturbation theory in gaussian normal coordinates happens to be more tractable.

\subsection{Next approximation in $\kappa$: unbroken $\tau$ symmetry}
Let us find the modifications of kink profiles and the shift of critical point under gravity influence. In the unbroken phase (zero order in $\mu$) the expansion in $\kappa$ reads,
\be
\tilde\Phi = M \sum_{n=0}^\infty \kappa^n \Phi_n,\quad \rho = \sum_{n=1}^\infty \kappa^n \rho_n . \label{phiexp}
\ee
In order to simplify the asymptotic behavior and analytic structure we introduce also the coupling dependence into the argument of iterated functions similar to Eq.\eqref{zeroap}),$ \beta \rightarrow \beta (\kappa) $ with the expansion,
\be
\frac{1}{\beta^2(\kappa)} = \sum_{n=0}^\infty \kappa^n \left(\frac{1}{\beta^2}\right)_n ;\quad \left(\frac{1}{\beta^2}\right)_0 = 1. \label{betaexp}
\ee
After rescaling $y = \tau / (\beta M), \tilde\Phi \rightarrow M \tilde\Phi$ the next-to-leading order for $\tilde\Phi$ obeys the equation,
\be
(\partial_\tau^2+2-6\Phi_{0}^2)\Phi_{1}=4\rho_{1}'\Phi_{0}' - 2 \kappa\left(\frac{1}{\beta^2}\right)_1 \Phi_0(1-\Phi_0^2)\equiv {\cal G}_1(\tau),\label{phione}
\ee
where the definitions \eqref{kink0} and \eqref{rho0} have been used.
Its real parity-odd solution can be found by integration of \eqref{phione},
\be \Phi_{1} =\frac{1}{\cosh^2\tau}\int^{\tau}_0 d\tau'
\cosh^4\tau'\int^{\tau'}_{-\infty} d\tau''\frac{1}{\cosh^2\tau''} {\cal G}_1(\tau). \ee
It decreases at infinity for $\left(\frac{1}{\beta^2}\right)_1  = 4/3$ and looks as follows,
\be
\Phi_{1}=-\frac{2}{9}\frac{\sinh{\tau}}{\cosh^3{\tau}}. \label{phi1}
\ee
 Therefrom the appropriately  iterated function $\tilde\Phi(y)$ can be represented as,
\be
\tilde\Phi(y) = M\tanh{\beta M y}\left(1 -\kappa\frac{2}{9 \cosh^2{\beta M y}}\right) + {\cal O}(\kappa^2);\quad \beta = 1 - \frac23\kappa . \label{phi12}
\ee
 The second approximation of conformal factor $\rho_{2}'$ derived directly from  \eqref{phi1} obeys the equation,
\be
\rho_{2}''= 2\tilde\Phi_{0}'\tilde\Phi_{1}',
\ee
which can be integrated to,
\be
\rho_{2}'=- \frac{2M}{135}\tanh{My}\left(38+ \frac{19}{\cosh^2{My}} + \frac{18}{\cosh^4{My}}\right) .
\ee
Accordingly the iterated result could be assembled in,
\ba
\rho (\tau) &=& \frac23 \kappa(1 -\frac{8}{45}\kappa) \log\cosh{\tau} + \frac{1}{6}\kappa\left(1 - \frac{26}{45}\kappa\right) \no&& - \frac{1}{6}\kappa\left(1 - \frac{8}{45}\kappa\right)\frac{1}{\cosh^2{\tau}} + \frac{1}{15} M\kappa^2\frac{1}{\cosh^4{\tau}} +
{\cal O}(\kappa^3) ,\no
&=& - \frac13 \kappa(1 -\frac{8}{45}\kappa) \log(1- \tanh^2{\tau}) \no&&+ \frac{\kappa}{6}(1 -\frac{44}{45}\kappa)  \tanh^2{\tau} + \frac{\kappa^2}{15} \tanh^4{\tau}+ {\cal O}(\kappa^3) ,\label{rho12} \ea
where the first expansion is ordered in accordance to its decreasing at large $y$ and the second one characterizes better the vicinity of $y = 0$ where the normalization $\rho(0) = 0$ is employed.
\subsection{Next approximation in $\kappa$: broken $\tau$ symmetry phase}

Above the phase transition point one discovers nontrivial solutions for $\tilde H (\tau)$ which satisfy the properly normalized Eq. \eqref{39} .
When a weak gravity is present then all functions and constants are taken depending on $\kappa$,
\ba &&\tilde H(\tau) = M\sum^\infty_{n,m=0}\kappa^n \Bigl(\frac{\mu}{M}\Bigr)^{2m+1}H_{n,m}(\tau);\quad
 \tilde\Phi(\tau) = M\sum^\infty_{n,m=0}\kappa^n \Bigl(\frac{\mu}{M}\Bigr)^{2m}\Phi_{n,m}(\tau);\quad \Phi_{n,0}\equiv  \Phi_{n}, \no&& \rho(\tau) = \kappa \sum^\infty_{n,m=0}\kappa^n \Bigl(\frac{\mu}{M}\Bigr)^{2m}\rho_{n+1,m}(\tau);\quad \rho_{n,0}\equiv  \rho_{n}, \no&&\Delta_H
  = \Delta_{H,c}(\kappa) + \frac12 \mu^2 ,\ea
   as well as,
\be
\frac{1}{\beta^2} = \sum^\infty_{n,m=0}\kappa^n \Bigl(\frac{\mu}{M}\Bigr)^{2m}
\Bigl(\frac{1}{\beta^2}\Bigr)_{n,m};\quad \Bigl(\frac{1}{\beta^2}\Bigr)_{0,0} = 1;\quad \Bigl(\frac{1}{\beta^2}\Bigr)_{0,1} = 1;\quad \Bigl(\frac{1}{\beta^2}\Bigr)_{1,0} =\frac43 .
\ee
The position of the critical point $\mu = 0$ is generically shifted,
\be
\Delta_{H,c}(\kappa) =\frac12 M^2 \sum^\infty_{n=0}\kappa^n  \Delta_H^{n} = \frac12 M^2 \left(1 - \frac{44}{27}\kappa\right) + {\cal O}(\kappa^2), \label{critshift}
\ee
which can be established from the consistency of integrated EoM. Indeed,
in the leading approximation against its normalization scale $\mu$ the function $\tilde H (\tau)$ satisfies the equation,
\ba&&
(\partial_\tau^2+1-2\Phi_{0,0}^2)H_{1,0}=\\&& -\kappa\left(\Delta^{1}_H + \left(\frac{1}{\beta^2}\right)_{1,0}\right) H_{0,0} +  4\rho_{1}' H_{0,0}' + 2 \kappa\left(\frac{1}{\beta^2}\right)_{1,0} H_{0,0}\Phi_{0,0}^2 + 4 H_{0,0}\Phi_{0,0}\Phi_{1,0}\equiv {\cal F}_1(\tau).\nonumber \label{Hone}
\ea
Its solution can be found by integration of \eqref{Hone},
\be
 H_{1,0} =\frac{1}{\cosh\tau}\left[C^H_{1,0} + \int^{\tau}_0 d\tau'
\cosh^2\tau'\int^{\tau'}_{0} d\tau''\frac{1}{\cosh\tau''} {\cal F}_1(\tau)\right].
\ee
and it is given by,
\be
 H_{1,0} = \frac{2}{27 \cosh\tau} \left(C^H_{1,0} - 2\log\cosh\tau + 3\tanh^2\tau\right) ,\label{h10}
\ee
provided that \eqref{critshift} holds. The integration constant $C^H_{1,0}$ is not fixed at this order in $\kappa, \mu$.

Mixed orders in $\kappa$ and $\mu^2/ M^2$ practically irrelevant as in realistic models $\kappa \sim 10^{-15}$ and $\mu^2/ M^2 \sim  10^{-3}$ (see \cite{aags2} and the Sect. 7). Correspondingly, $\kappa\mu^2/ M^2 \ll \kappa \ll \mu^2/ M^2$. Therefore the overlapping of classical solutions \eqref{zeroap}, \eqref{rho1} with solutions \eqref{phi12} \eqref{rho12}, \eqref{h10} provides our calculations with required precision in the case when the perturbation expansion works well. The latter seems to be flawless for classical EoM.

\section{Field fluctuations around the classical solutions}
\subsection{Quadratic action and infinitesimal diffeomorphisms}
We consider small localized deviations of the fields
from the average background values and find the action-square corresponding to them.

Action \eqref{1} is invariant under diffeomorphisms. Infinitesimal diffeomorphisms correspond to the Lie derivative along an arbitrary vector field $ \tilde \zeta ^A (X) $, defining the coordinate transformation $ X \to \tilde X = X +\tilde \zeta \left (X \right) $ .

Let us introduce the fluctuations of the metric $ h_ {AB} \left (X \right) $ and the scalar fields $ \phi \left (X \right) $ and $\chi \left (X \right) $
 on the background solutions of the equations of motion,
    \ba
&&g_{AB} \left( X \right) = A^2 \left( z \right)\left( {\eta _{AB}  + h_{AB} \left( X \right)} \right);\no&& \Phi \left( X \right) = \Phi \left( z \right) + \phi \left( X \right);\quad H \left( X \right) = H \left( z \right) + \chi \left( X \right) \label{conf} . \ea

Since 4D Poincare symmetry is not broken, we select the corresponding 4D part of the metric $ h_ {\mu \nu} $ and introduce the notation for gravivectors $ h_ {5 \mu} \equiv v_ \mu $ and graviscalars $ h_ {55} \equiv S $. By rescaling the vector fluctuations $ \tilde \zeta _ \mu = A ^ 2 \zeta _ \mu $ and the scalar ones $ \tilde \zeta _5 = A \zeta _5 $, we obtain the following gauge transformations
 in the first order of $ \zeta^ A(X) $ ,  \[
h_{\mu \nu }  \to h_{\mu \nu }  - \left( {\zeta _{\mu ,\nu }  + \zeta _{\nu ,\mu }  - \frac{{2A'}}{{A^2 }}\eta _{\mu \nu } \zeta _5 } \right) ,\qquad v_\mu   \to v_\mu   - \left( {\frac{1}{A}\zeta _{5,\mu }  + \zeta '_\mu  } \right) ,\]\be S \to S - \frac{2}{A}\zeta '_5 ,\qquad \phi  \to \phi  + \zeta _5 \frac{{\Phi '}}{A} ,\qquad \chi  \to \chi  + \zeta _5 \frac{{H '}}{A} ,\label{10} \ee
 with an accuracy of order $O\left( {\zeta ^2 ,h^2 ,h\zeta } \right)$. Herein "$,$"\ denotes a partial derivative.

Now expand the action to quadratic order in fluctuations. The full
action after this procedure is a sum,
 \be {\cal L}_{(2)}= {\cal L}_{h}+{\cal L}_{\phi,\chi}+{\cal L}_S+{\cal L}_{V}, \ee where \ba \sqrt {\left| g \right|}{\cal L}_{h}&\equiv&\ -\frac12 M^3_\ast A^3\ \Bigl\lbrace -\frac{1}{4}\ h_{\alpha\beta,\nu}h^{\alpha\beta,\nu} -\frac{1}{2}\ h^{\alpha\beta}_{,\beta}h_{,\alpha}+\frac{1}{2}\ h^{\alpha\nu}_{,\alpha}h^{\beta}_{\nu,\beta}+\frac{1}{4}\ h_{,\alpha}h^{,\alpha} \no &&+ \frac{1}{4}h'_{\mu\nu}h'^{\mu\nu}-\frac{1}{4}h'^2 \Bigr\rbrace , \ea
 \ba \sqrt {\left| g \right|} {\cal L}_{\phi,\chi}&\equiv&\frac{1}{2} A^3 (\phi_{,\mu}\phi^{,\mu}- \phi'^2+ \chi_{,\mu} \chi^{,\mu}- (\chi')^2) - \frac{1}{2}\ A^5 \Big( \frac{\partial^2 V}{\partial\Phi^2} \phi^2 + 2 \frac{\partial^2 V}{\partial\Phi\partial H} \phi \chi+ \frac{\partial^2 V}{\partial H^2} \chi^2\Big)\no &&+ \frac{1}{2}   A^3 h'(\Phi'\phi+ H'\chi) , \ea
 \ba
 \sqrt {\left| g \right|}{\cal L}_{S}&\equiv& \frac{1}{4}\Big(-A^5VS^2+  S \Bigl( M^3_\ast A^3 \left(h^{\mu\nu}_{,\mu\nu}-h^{,\mu}_{,\mu} \right) + M^3_\ast \left( A^3\right)'h'\no&&+ 2 \big(A^3(\Phi'\phi+ H'\chi)\big)' - 4A^3(\Phi'\phi'+ H'\chi')\Bigr)\Big),\label{S}
 \ea
\ba \sqrt {\left| g \right|}{\cal L}_{V}&\equiv& - \frac{1}{8}M^3_\ast
A^3 v_{\mu\nu}v^{\mu\nu}+ \frac12 v^\mu \Big[ - M^3_\ast A^3
\left(h_{\mu\nu}^{,\nu}-h_{,\mu} \right)'\no&&+ 2A^3(\Phi'\phi_{,\mu}+ H'\chi_{,\mu}) +
M^3_\ast\Bigl( A^3\Bigr)'S_{,\mu}\Big] \label{V} ,
 \ea
 where $v_{\mu\nu}=v_{\mu,\nu}-v_{\nu,\mu}$, $h=h_{\mu\nu}\eta^{\mu\nu}$. Transformations (\ref{10}) allow to eliminate gauge degrees of freedom.

\subsection{Disentangling the physical degrees of freedom}
A physical sector can be determined after the separation of
different spin components of the field $ h_ {\mu \nu} $ in the
system. It  can be accomplished by description of ten components of
4-dim metric in terms of the traceless-transverse tensor, vector and
scalar components \cite{rev15,bar},
\ba h_ {\mu \nu} = b_ {\mu \nu}
+ F_ {\mu, \nu} + F_ {\nu, \mu} + E_ {, \mu \nu} + \eta_ {\mu \nu}
\psi, \label {deco} \ea where $ b_ {\mu \nu } $ and $ F_ \mu $ obey
the relation $ b_ {\mu \nu} ^ {, \mu} = b = 0 = F_ {\mu} ^ {, \mu}
$. Obviously, the gravitational fields $ b_ {\mu \nu} $ are gauge
invariant and thereby describe graviton fields in the 4-dim space.
Let's expand the gauge parameter $ \zeta_ \mu $ and  vector fields $
v_ \mu $ into the transverse and longitudinal parts,
\be \zeta_ \mu
= \zeta_ \mu ^ \perp + \partial_ \mu C, \qquad \partial ^ \mu \zeta_
\mu ^ \perp = 0; \qquad v_ \mu = v_ \mu ^ \perp + \partial_ \mu
\eta, \qquad \partial ^ \mu v_ \mu ^ \perp = 0. \ee Then the vector
fields are transformed as follows,
\be F_ \mu \rightarrow F_ \mu -
\zeta_ \mu ^ \perp, \quad v_ \mu ^ \perp \rightarrow v_ \mu ^ \perp
- {\zeta'_ \mu } ^ \perp, \ee i.e. the expression $ F'_ \mu - v_ \mu
^ \perp $ is gauge invariant. In turn, the scalar components $ \eta,
E, \psi, S, \phi $ change under gauge transformations in the
following way,
\ba & & \eta \rightarrow \eta - \frac {1} {A} \zeta_5
- C ' ; \qquad E \rightarrow E - 2C, \no & & \psi \rightarrow \psi +
\frac {2A '} {A ^ 2} \zeta_5, \qquad S \rightarrow S - \frac {2} {A}
\zeta'_5, \qquad \phi \rightarrow \phi + \frac {\Phi '} {A} \zeta_5
, \qquad \chi \rightarrow \chi + \frac {H '} {A} \zeta_5.
\ea Therefrom  we can find four independent gauge invariants,
 \ba \frac12 E' - \eta -  \frac{A}{2A'} \psi;\quad -\psi + \frac{2A'}{A \Phi'} \phi;\quad \frac12 A S + \left(\frac{A}{\Phi'} \phi\right)';\quad H' \phi - \Phi' \chi . \ea
Using the parametrization (\ref{deco}) we can calculate components
of the quadratic action, \ba && h \equiv h^\mu_\mu = \square E + 4
\psi;\quad h^{\alpha\beta}_{,\beta} = \square ( F^\alpha
+E^{,\alpha}) + \psi^{,\alpha};\quad h^{\alpha\beta}_{,\alpha\beta}
= \square^2 E + \square \psi;\no &&
h^{\mu\nu}_{,\mu\nu}-h^{,\mu}_{,\mu} = - 3 \square \psi;\quad
h_{\mu\nu}^{,\nu}-h_{,\mu} = \square F_\mu - 3 \psi_{,\mu} . \ea
 Thus, the decomposition \eqref {deco} entails a partial separation of degrees of freedom in the lagrangian quadratic in fluctuations,
 \ba \sqrt{\left| g \right|} {\cal L}_{(2)} &=& \frac{1}{8} M^3_\ast A^3\ \Bigl\lbrace \ b_{\mu\nu,\sigma} b^{\mu\nu,\sigma} - (b')_{\mu\nu} (b')^{\mu\nu} -  f_{\mu\nu} f^{\mu\nu}\Bigr\rbrace\no &&+ \frac{3}{4} M^3_\ast A^3\ \Bigl\lbrace - \psi_{,\mu} \psi^{,\mu} + \psi_{,\mu} S^{,\mu} + 2 (\psi')^2 + 4 \frac{A'}{A} \psi' S\Bigr\rbrace\no &&+ \frac12 A^3\ \Bigl\lbrace \phi_{,\mu} \phi^{,\mu} - (\phi')^2 + \chi_{,\mu} \chi^{,\mu}- (\chi')^2 - \ A^2 \Big( \frac{\partial^2 V}{\partial\Phi^2} \phi^2 + 2 \frac{\partial^2 V}{\partial\Phi\partial H} \phi \chi+ \frac{\partial^2 V}{\partial H^2} \chi^2\Big)\no &&-\frac{1}{2}A^2 V(\Phi, H) S^2 + 4 \psi'(\Phi'  \phi +H' \chi) + S\Bigl(- \Phi' \phi' - H' \chi'+  A^2 \Big(\frac{\partial V}{\partial\Phi} \phi + \frac{\partial V}{\partial H} \chi\Big) \Bigr)\Bigr\rbrace\no &&+ \frac{3}{4} M^3_\ast A^3\ \square(E' - 2\eta) \Bigl(\frac{A'}{A} S + \psi' + \frac{2}{3M^3_\ast} (\Phi' \phi + H' \chi )\Bigr), \label{decoup} \ea
 where $ f_\mu \equiv F'_\mu  - v_\mu^\perp,\quad f_{\mu\nu} \equiv f_{\mu,\nu} - f_{\nu,\mu} $ .

We see that some redundant degrees of freedom exist , one of vectors
$ F'_ \mu, v_ \mu ^ \perp $ and one of scalars $ E ', \eta $. They
can be removed to provide $ v_ \mu = 0 $. Obviously, in the
quadratic approximation graviton, gravivector and graviscalar are
decoupled from each other. From the last line it follows that the
scalar $ E '$ is a Lagrange multiplier and generates a
gauge-invariant constraint, \ba \label{cond1} \frac{A'}{A} S + \psi
'= - \frac {2 } {3M ^ 3_ \ast}(\Phi' \phi + H' \chi ) . \ea Thus taking this
constraint into account   only
two independent scalar fields  remain.
\section{Scalar field action in gauge invariant variables}

The further analysis of the scalar spectrum is convenient to perform in gauge invariant variables. Let us perform the following rotation in $(\phi,\chi)$ sector:
\ba \phi=\check{\phi}\cos{\theta}+\check{\chi}\sin{\theta},\quad
\chi=-\check{\phi}\sin{\theta}+\check{\chi}\cos{\theta}\no
\cos{\theta}=\frac{\Phi'}{\mathcal{R}},\quad\sin{\theta}=\frac{H'}{\mathcal{R}},\quad
\mathcal{R}^2=(\Phi')^2+(H')^2 \ea

While $\check{\chi}$ is gauge invariant $\check{\phi}$ is not. We can exclude redundant gauge invariance introducing  three gauge invariant variables:
\ba \check{\psi}=\psi-\frac{2A'}{A\mathcal{R}}\check{\phi},\quad
\check{S}=S+\frac{2}{\mathcal{R}}\check{\phi}'-\frac{2A}{\mathcal{R}^2}
\left(\frac{\mathcal{R}}{A}\right)'\check{\phi},\quad
\check{\eta}=E'-2\eta-\frac{2}{\mathcal{R}}\check{\phi} .
\ea

Accordingly the scalar part of the lagrangian quadratic in fluctuations takes the form:
 \ba \sqrt{\left| g \right|} {\cal L}_{(2),scal} &=& \frac{3}{4} M^3_\ast A^3\ \Bigl\lbrace - \check{\psi}_{,\mu} \check{\psi}^{,\mu} + \check{\psi}_{,\mu} \check{S}^{,\mu} + 2 (\check{\psi}')^2 + 4 \frac{A'}{A} \check{\psi}' \check{S}\Bigr\rbrace+ \frac12 A^3\ \Bigl\lbrace \check{\chi}_{,\mu} \check{\chi}^{,\mu}- (\check{\chi}')^2 -\no && -\Big[(\theta')^2+\frac{A^2}{\mathcal{R}^2} \Big( \frac{\partial^2 V}{\partial\Phi^2} (H')^2 - 2 \frac{\partial^2 V}{\partial\Phi\partial H} \Phi'H'+ \frac{\partial^2 V}{\partial H^2} (\Phi')^2\Big)\Big]\check{\chi}^2\Bigr\rbrace+A^3\mathcal{R}\theta'\check{S}\check{\chi}-\no &&
 -\frac{1}{4}A^5 V(\Phi, H) \check{S}^2+ \frac{3}{4} M^3_\ast A^3\ \square\check{\eta}\Bigl(\frac{A'}{A} \check{S}
  + \check{\psi}'\Bigr)\Bigr) .
 \ea
where $\theta'=(\arctan{\frac{H'}{\Phi'}})'=(H''\Phi'-\Phi''H')/\mathcal{R}^2$

From the last line it follows that the
scalar field $\check{\eta}$ is a gauge invariant Lagrange multiplier and generates a gauge invariant
constraint, \ba \label{cond2} \frac{A'}{A} \check{S} + \check{\psi}'= 0. \ea
Thus after taking this constraint into account  only two independent scalar fields remain and
the scalar action takes the following form,
\ba
\sqrt{\left| g \right|}{\cal L}_{(2),scal}=
\frac{A^5\mathcal{R}^2}{8(A')^2}\left\{\partial_{\mu}\check{\psi}\partial^{\mu}\check{\psi}
-(\partial_z\check{\psi})^2\right\}-
 \frac{A^4}{A'}\mathcal{R}\theta'(\partial_z\check{\psi})\check{\chi}+\no
 +\frac{A^3}{2}\left\{\partial_{\mu}\check{\chi}\partial_{\mu}\check{\chi}
 -(\partial_z\check{\chi})^2-
 \left((\theta')^2+\frac{A^2}{\mathcal{R}^2}\begin{pmatrix}H'\\
 -\Phi'\end{pmatrix}^{\dag}\partial^2V\begin{pmatrix}H'\\-\Phi'\end{pmatrix}\right)\check{\chi}^2\right\}
 \ea
To normalize kinetic terms the fields should be redefined $\hat{\chi}=A^{3/2}\check{\chi}$, $\hat{\psi}=\Omega\check{\psi}$, where $\Omega=A^{5/2}\mathcal{R}/2A'$.
\ba
\sqrt{\left| g \right|} {\cal L}_{(2),scal}=\frac{1}{2}\left\{\partial_{\mu}\hat{\psi}\partial^{\mu}\hat{\psi}
-(\partial_z\hat{\psi})^2-\frac{\Omega''}{\Omega}\hat{\psi}^2\right\}
-2\theta'\hat{\chi}\left(\partial_z-\frac{\Omega'}{\Omega}\right)\hat{\psi}\no
+\frac{1}{2}\left\{\partial_{\mu}\hat{\chi}\partial^{\mu}\hat{\chi}
-(\partial_z\hat{\chi})^2-\frac{(A^{3/2})''}{A^{3/2}}-\left((\theta')^2
+\frac{A^2}{\mathcal{R}^2}\begin{pmatrix}H'\\-\Phi'\end{pmatrix}^{\dag}
\partial^2V\begin{pmatrix}H'\\-\Phi'\end{pmatrix}\right)\hat{\chi}^2\right\}
\ea
\section{Fluctuations in different phases and at critical point}
\subsection{Fluctuations around  a $\tau$ symmetric background}
When $H(z) = 0$ the two scalar sectors decouple because $\theta = 0$.  The operator which describes the branon mass spectrum, \be\hat m^2_\psi = -\partial_z^2 + \frac{\Omega''}{\Omega} = \Big(\partial_z + \frac{\Omega'}{\Omega}\Big)\Big( -\partial_z + \frac{\Omega'}{\Omega}\Big),\ee  is positive on functions $\hat{\psi}(z)$ normalizable along the fifth dimension $z$. Indeed, the possible zero mode is singular $\hat{\psi}(z) \sim \Omega \sim 1/z\Big|_{z\rightarrow 0}$. It corresponds to the centrifugal barrier in the potential $ \Omega''/\Omega$ at the origin \cite{rev19}. Thus in the presence of gravity there is no a (normalizable) Goldstone zero-mode related to spontaneous breaking of translational symmetry. The cause is evident: the corresponding brane fluctuation represents, in fact, a gauge transformation \eqref{10} and does not appear in the invariant part of the spectrum. One could say that in the presence of gravity induced by a brane the latter becomes  more rigid as only massive fluctuations are possible around it. Of course, the very gauge transformation \eqref{10} leaves invariant only the quadratic action and  thereby a track of Goldstone mode may have influence on higher order vertices of interaction between gravity and scalar fields. This option is beyond the scope of the present investigation.

As to the possible localized states with positive $m^2_\psi > 0$ they may exist with masses of order $M$. However for the action \eqref{36} they happen to be unstable resonances as it will be evident from the spectral problem formulated in gaussian normal coordinates.

The fluctuations of the second, mass generating field $H(x)$ do not develop any centrifugal barrier and as $<H> = 0$ their mass spectrum is described by the operator,
\be
\hat m^2_\chi = - \partial_z^2 + \frac{(A^{3/2})''}{A^{3/2}}
+A^2
\frac{\partial^2 V(\Phi,H))}{(\partial H)^2}\Big|_{H = 0} \equiv - \partial_z^2 + {\cal V}(z). \label{higg}
\ee
Its potential is not singular and for background solutions delivering a minimum this operator must be positive. For the minimal potential with quartic self-interaction \eqref{36} (in terms of the rescaled variables \eqref{variab}) one can come to more quantitative conclusions. Indeed, for gravity switched off the background $\tilde\Phi(z) = \Phi_0(z)$ (pay attention to $y \rightarrow z$!) is defined by \eqref{kink0} . Accordingly the mass spectrum operator receives the potential
\be
{\cal V}(z) = - 2\Delta_H + 2 \Phi_0^2 = (M^2 - 2\Delta_H) + M^2\Big(1 - \frac{2}{\cosh^2 Mz}\Big).
\ee
The only localized state of the mass operator  $\hat m^2_\chi$ is $\hat\chi \rightarrow \chi_0 \simeq 1/\cosh(Mz)$ with the corresponding mass $m^2_0 = M^2 - 2\Delta_H$ as expected. Thus in the unbroken phase with $M^2 > 2\Delta_H$ the lightest scalar fluctuation in $\chi$ channel possesses a positive mass and the system is stable.
In the critical point,  $M^2 = 2\Delta_H$, a lightest fluctuation is massless and for $M^2 < 2\Delta_H \leq 2M^2$ the localized state $\chi_0$ represents a tachyon and brings instability providing a saddle point. Instead the solution \eqref{zeroap} provides a true minimum (see \cite{aags1}).

Qualitatively the spectrum pattern in the gravity background remains similar. But the derivation of localized eigenfunctions uniformly in coordinate $z$ encounters certain difficulties as explained in the Subsection \ref{confgauss} and therefore it will be done in gaussian normal coordinates.
\subsection{Fluctuations in gaussian normal coordinates}
To simplify analytical calculations let us represent the quadratic
action for scalar fields in the gaussian normal coordinates $x_\mu,
y$, \be ds^2  = A^2 \left( z \right)\left( {dx_\mu  dx^\mu   - dz^2
} \right) = \exp \left( { - 2\rho \left( y \right)} \right)dx_\mu
dx^\mu   - dy^2 .\ee We remind the formulas for the
transition,
    \[
z = \int {\exp \rho \left( y \right)dy} ,\quad A\left( z \right) =
\exp \left( { - \rho \left( y \right)} \right) .
\]
Below the prime denotes differentiation with respect to $y$. Further on we focus on the minimal potential with quartic self-interaction \eqref{36} in terms of the rescaled variables \eqref{variab}).
To simplify the form of the action let us introduce $\widetilde{\mathcal{R}}=\exp(\rho)\mathcal{R}$ and in addition
redefine the fields in order to normalize kinetic term, $\hat{\psi}=\exp(-\rho/2)\tilde{\psi}$,$\hat{\chi}=\exp(-\rho/2)\tilde{\chi}$.
\ba
&&\!\!\!\!\!\!\!\!\!\!\!S_{(2),scal}=\int d^4xdy\left[\frac{1}{2}\partial_{\mu}\tilde{\psi}\partial^{\mu}\tilde{\psi}
+\frac{1}{2}\partial_{\mu}\tilde{\chi}\partial^{\mu}\tilde{\chi}
-2\exp(-2\rho)\theta'\tilde{\chi}\left(\partial_y+\rho'+\frac{\rho''}{\rho'}
-\frac{\widetilde{\mathcal{R}}'}{\widetilde{\mathcal{R}}}\right)\tilde{\psi}-\right.\\
&&\!\!\!\!\!\!\!\!\!\!\!-\frac{1}{2}\exp(-2\rho)\tilde{\psi}\left\{\left(-\partial_y+\frac{\rho''}{\rho'}
-\frac{\widetilde{\mathcal{R}}'}{\widetilde{\mathcal{R}}}\right)\left(\partial_y
+\frac{\rho''}{\rho'}-\frac{\widetilde{\mathcal{R}}'}{\widetilde{\mathcal{R}}}\right)
+2\rho'\partial_y+3(\rho')^2+3\rho''-4\rho'
\frac{\widetilde{\mathcal{R}}'}{\widetilde{\mathcal{R}}}\right\}\tilde{\psi}-\no
&&\!\!\!\!\!\!\!\!\!\!\!\left.-\frac{1}{2}\exp(-2\rho)\tilde{\chi}\left\{-\partial_y^2+(\theta')^2
+\frac{1}{\widetilde{\mathcal{R}^2}}\begin{pmatrix}\tilde H'\\
-\tilde\Phi'\end{pmatrix}^{\dag}\partial^2V\begin{pmatrix}\tilde H'\\
-\tilde\Phi'\end{pmatrix}+2\rho'\partial_y+3(\rho')^2-\rho''\right\}\tilde{\chi}\right].\nonumber
\ea
where the second variation of the  field potential reads,
\ba
\partial^2V = \left(\begin{array}{cc}- 2M^2 +6 \tilde\Phi^2 + 2 \tilde H^2 & 4 \tilde\Phi \tilde H\\
4\tilde\Phi\tilde H& -2\Delta_H + 2\tilde\Phi^2 +6 \tilde H^2\end{array}\right) .
\ea
Let us perform the mass spectrum expansion,
\ba
&&\tilde{\psi}(X)=\exp(\rho)\sum_m \Psi^{(m)}(x)\psi_m(y),\quad\tilde{\chi}(X)=\exp(\rho)\sum_m\Psi^{(m)}(x)\chi_m(y),\no
&&\partial_{\mu}\partial^{\mu}\Psi^{(m)}=-m^2\Psi^{(m)},
\ea
where the factor $\exp(\rho)$ is introduced to  eliminate first derivatives in the equations.
We obtain the following equations,
\ba
\left(-\partial_y+\frac{\rho''}{\rho'}-\frac{\widetilde{\mathcal{R}}'}{\widetilde{\mathcal{R}}}
+2\rho'\right)\left(\partial_y+\frac{\rho''}{\rho'}
-\frac{\widetilde{\mathcal{R}}'}{\widetilde{\mathcal{R}}}+2\rho'\right)\psi_m-\no
-2\theta'\left(\partial_y-\frac{\rho''}{\rho'}
+\frac{\widetilde{\mathcal{R}}'}{\widetilde{\mathcal{R}}}
-2\rho'+\frac{\theta''}{\theta'}\right)\chi_m=\exp(2\rho)m^2\psi_m,\\
\left(-\partial_y^2+(\theta')^2+\frac{1}{\widetilde{\mathcal{R}^2}}\begin{pmatrix}\tilde H'\\
-\tilde\Phi'\end{pmatrix}^{\dag}\partial^2V\begin{pmatrix}\tilde H'\\-\tilde\Phi'\end{pmatrix}+4(\rho')^2
-2\rho''\right)\chi_m+\no
+2\theta'\left(\partial_y+\frac{\rho''}{\rho'}
-\frac{\widetilde{\mathcal{R}}'}{\widetilde{\mathcal{R}}}
+2\rho'\right)\psi_m=\exp(2\rho)m^2\chi_m . \label{flucspec}
\ea
This is a coupled channel equation of second order in derivative and with the spectral parameter $m^2$ as being a coupling constant of a part of potential (a non-derivative piece). The latter part is essentially negative for all $m^2 >0$. Then as the exponent $\rho(y)$ is positive and growing at very large $y$ it becomes evident that the mass term in the potential makes it unbounded below. Thus any eigenfunction of the spectral problem \eqref{flucspec} is at best a resonance state though it could be quasilocalized in a finite volume around a local minimum of the potential. In \cite{aags2} the probability for quantum tunneling of quasilocalized light resonances with masses $m\ll M$ was estimated as $\sim \exp\{-\frac{3}{\kappa}\ln\frac{2M}{m}\}$ which for phenomenologically acceptable values of $\kappa \sim 10^{-15}$ and $M/m \gtrsim 30$ means an enormous suppression. Moreover in the perturbation theory the decay does not occur as the turning point to an unbounded potential energy is situated at $y\sim 1/\kappa$.  Therefore one can calculate the localization of resonances  following the perturbation schemes.

In the limit $\kappa\longrightarrow 0$ we obtain,
\ba
\left(-\partial_y+\frac{\rho_1''}{\rho_1'}-\frac{\widetilde{\mathcal{R}}'}{\widetilde{\mathcal{R}}}
\right)\left(\partial_y+\frac{\rho_1''}{\rho_1'}
-\frac{\widetilde{\mathcal{R}}'}{\widetilde{\mathcal{R}}}\right)\psi_m-2\theta'
\left(\partial_y-\frac{\rho_1''}{\rho_1'}+\frac{\widetilde{\mathcal{R}}'}{\widetilde{\mathcal{R}}}
+\frac{\theta''}{\theta'}\right)\chi_m=m^2\psi_m,\\
\left(-\partial_y^2+(\theta')^2+\frac{1}{\widetilde{\mathcal{R}^2}}\begin{pmatrix}\tilde H'\\
-\tilde\Phi'\end{pmatrix}^{\dag}\partial^2V\begin{pmatrix}\tilde H'\\-\tilde\Phi'\end{pmatrix}\right)\chi_m
+2\theta'\left(\partial_y+\frac{\rho_1''}{\rho_1'}
-\frac{\widetilde{\mathcal{R}}'}{\widetilde{\mathcal{R}}}\right)\psi_m=m^2\chi_m
\ea
where $\rho_1$ is first order of $\kappa$.
\subsection{Phase transition point in the presence of gravity}

In the unbroken phase $\tilde H(y) = 0$ and the equation on $\chi$ takes the form,
\be
\Bigl[-\partial_\tau^2+ \frac{1}{\beta^2 M^2}e^{-2\rho}\Bigl(-2\Delta_H+2\Phi^2\Bigr) + 4 (\rho')^2 - 2 \rho''\Bigr]\chi_m=
\frac{m^2}{M^2\beta^2}e^{2\rho}\chi_m , \label{gauchi}
\ee
where the variable $\tau = \beta M y$ is employed and the derivative is defined against it.

Let us perform the perturbative expansion in $\kappa$, \be\chi_m = \sum_{n=0} \kappa^n \chi_{m,n},\quad \Delta_{H,c} = \frac12 M^2 \sum_{n=1} \kappa^n \Delta^n_H;\quad
m^2 = \sum_{n=1} \kappa^n (m^2)_n\ee and use also the expansions \eqref{phiexp} and \eqref{betaexp}. The limit of turned off gravity is smooth and the differential operator on the left-hand side of \eqref{gauchi} can be factorized,
\be
\Bigl[\frac{M^2-2 \Delta_H}{M^2}+(-\partial_\tau+\tanh\tau)(\partial_\tau+
\tanh\tau)\Bigr]\chi_{m,0} =\frac{(m^2)_0}{M^2}\chi_{m,0} ,
\ee
which  corresponds to $\Delta^0_H = 1$ for zero scalar mass (phase transition point).
In general, for $M^2-2\Delta_H >0$ one finds one localized state with positive $m^2$,
\be
\chi=\frac{1}{\cosh\tau}+O(\kappa),\quad m^2=M^2-2\Delta_H+O(\kappa),
\ee
as it was already established in the previous Subsection.

Let us now examine the phase transition point where $m^2 = 0$ and calculate the next approximation of $\kappa$:
\ba
& 0=\Bigl[-\partial_\tau^2 +1 - \frac{2}{\cosh^2\tau}
\Bigr]\chi_1+\no
&+\Bigr[\Bigl(\frac{1}{\beta^2}\Bigr)_1\Bigl(1 - \frac{2}{\cosh^2\tau}\Bigr)-
\Delta_H^{1}+4\Phi_0\Phi_1-2\rho_1''\Bigl]\chi_0 . \label{zerochi}
\ea
At critical point $\Delta_H=\Delta_{H,c}=\frac{1}{2}M^2(1 -\frac{44}{27}\kappa+O(\kappa^2))$ exactly as it has been obtained in \eqref{critshift}. Accordingly there exists a normalizable solution of \eqref{zerochi} which is a
zero-mode corresponding to the second-order phase transition. In this case the corresponding first correction of $\chi$  takes the form,
\be
\chi_1=\frac{1}{9}\frac{1}{\cosh\tau}\Bigl[\frac{1}{\cosh^2\tau}-\frac{40}{3}\ln(2\cosh\tau)
+ \frac{38}{3} + C_1\Bigr],
\ee
where for the constant $C_1 = 0$ this correction is orthogonal to $\chi_0$. Thus in the scalar sector not mixed with branon (gravity) fluctuations the localization of massless state occurs in the presence of gravity. It can also be shown that for $\Delta_H < \Delta_{H,c}$ the  quasilocalization of light states in this sector takes place.

When $\Delta_H>\Delta_{H,c}$ the squared mass becomes negative signalling the instability of the unbroken phase.
In broken phase mixing terms are nonzero and one has to study spectrum by perturbation theory near
 critical point. The calculations are not presented in this paper because of their high complexity but to the leading order in $\kappa$ they provide the
  same mass for light scalar state as in the model \cite{aags1} without gravity, namely, $m^2=2\mu^2+O(\mu^4/M^2)$. This state is associated with the fermion mass generation (Sect. 2) and substitutes the Higgs field of the Standard Model.
 \section{Conclusions: consistency of scales and of gravitational coupling with modern data}
To consider phenomenological implications we have to study interaction of the scalar matter with fermions,
\be
\mathcal{L}_{f}=\bar{\Psi}(i\partial\!\!\!/-g_{K}\tau_3\Phi-g_{H}\tau_1H)\Psi,
\ee
where in general we can introduce different Yukawa constants for different fermions of the Standard Model (SM). The localization profile depends on the first coupling $g_K$,
\be
\psi_0=\exp\Bigl(-g_K\int^ydy'\Phi(y')\Bigr)=\frac{1}{\cosh^{\alpha}{M\beta y}},\quad \alpha=\frac{g_K}{\beta}=g_K+O\Bigl(\frac{\mu^2}{M^2}\Bigr).
\ee
Correspondingly in the leading order in $\mu$ and $\kappa$ the fermion mass is described by
\be
m_f=\frac{\int_{-\infty}^{+\infty}\psi_0(y)^2H(y)dy}{\int_{-\infty}^{+\infty}\psi_0(y)^2}
=g_{H}\mu\frac{\Gamma\Bigl(\alpha+\frac{1}{2}\Bigr)^2}{\Gamma\Bigl(\alpha\Bigr)\Gamma\Bigl(\alpha+1\Bigr)}.
\ee
As it was shown in Sections 3, 6 the scalar fluctuations have a single normalizable state associated with the fermion mass generation,
\be
\Phi=\Phi_0(y)+O\Bigl(\frac{\mu}{M}\Bigr),\quad H=H_0(y)+\chi_{0}(y)h(x)+O\Bigl(\frac{\mu^2}{M^2}\Bigr),\quad \chi_{0}=\frac{1}{\cosh{M\beta y}},
\ee
with the mass, $m_h=\sqrt{2}\mu\Bigl(1+O\Bigl(\frac{\mu^2}{M^2}\Bigr)\Bigr)$. For $\mu\ll M$ low energy four-dimensional Lagrangian including only the lightest states takes the following form,
\ba
\mathcal{L}_{low}&=&\frac{3\kappa M_{\ast}^3}{2M^3}\int_{-\infty}^{+\infty}\chi_0(y)^2dy\cdot
\Bigl(\partial_{\mu}h\partial^{\mu}h-m_h^2h^2\Bigr)+2\int_{-\infty}^{\infty}\psi_0(y)^2dy\cdot\bar{\psi}
\Bigl(i\partial\!\!\!/-m_f\Bigr)\psi-\no
&&-2g_H\int_{-\infty}^{+\infty}\psi_0(y)^2\chi_0(y)dy\cdot\bar{\psi}h\psi .
\ea
After normalization,
\be
h\rightarrow h\sqrt{\left.\frac{2}{3\kappa}\Bigl(\frac{M}{M_{\ast}}\Bigr)^3
\middle/\int_{-\infty}^{+\infty}\chi_0(y)^2dt\right.},\quad \psi\rightarrow \left.\psi\middle/\sqrt{2\int_{-\infty}^{+\infty}\psi_0(y)^2dy}\right. ,
\ee
we obtain the following Yukawa coupling constant between Higgs-like boson and fermion,
\be
g_f=\sqrt{\frac{2}{3\kappa}\Bigl(\frac{M}{M_{\ast}}\Bigr)^3}g_{H}
\frac{\int_{-\infty}^{+\infty}\psi_0(y)^2\chi_0(y)dy}{\sqrt{\int_{-\infty}^{+\infty}\chi_0(y)^2dt}
\int_{-\infty}^{+\infty}\psi_0(y)^2dy}=\sqrt{\frac{2}{3\kappa}\Bigl(\frac{M}{M_{\ast}}\Bigr)^3}
\frac{m_f}{m_h}.
\ee
We can compare it with similar couplings $\lambda, g_{t,SM}$ in the standard Higgs model.
We adopt the normalization of coupling constants in the Higgs potential of the Standard Model as follows,
\be
V_{SM}\big(h(x)\big) \equiv - m^2 h^2 + \lambda h^4,\quad \langle h \rangle = \frac{v}{\sqrt{2}} = \frac{m}{\sqrt{2\lambda}} .
\ee
The scale $v \simeq 246 GeV$ stands for the v.e.v of the Higgs field $h$ in the Standard Model \cite{PDG}.
 For the top quark channel dominating for the Higgs boson decay via one-loop mechanism one obtains,
\be
m_h=\sqrt{2\lambda} v,\quad m_t=\frac{1}{\sqrt{2}}g_{t,SM}\cdot v\Rightarrow g_{t, SM}=2\sqrt{\lambda}\frac{m_t}{m_h}.
\ee
Accordingly the relation between the Yukawa coupling constants is given by,
\be
\lambda\frac{g_t^2}{g_{t,SM}^2}=\frac{1}{6\kappa}\Bigl(\frac{M}{M_{\ast}}\Bigr)^3 . \label{yukawa}
\ee
Let us involve the gravity scales coming from reduction of five-dimensional Einstein-Hilbert action to the four-dimensional one \cite{aags2},
\be
M^3_\ast = k M^2_{P}, \label{planck}
\ee
which can be derived from the graviton kinetic action \eqref{decoup} when taking the wave function $b'_{\mu\nu}= 0$ for massless graviton. It determines the four-dimensional gravity scale, the Planck mass, $M_P \simeq 2.5\cdot 10^{18} GeV$ \cite{PDG}.

From the experimental bounds on the AdS curvature in extra dimension \cite{adelb} one can estimate the minimal value for the mass scales, $M_\ast, M$ as well as for the dimensional gravitational coupling $\kappa$. Indeed, if combining \eqref{planck}, \eqref{asymp} and \eqref{yukawa} one gets,
\be
M = \sqrt{3\sqrt{\lambda}\ k M_P \frac{g_t}{g_{t,SM}}} ;\quad \kappa = \frac{1}{2\sqrt{\lambda}}\frac{M}{M_P}\frac{g_{t,SM}}{g_t} .
\ee
The modern bound for the AdS curvature, $k > 0.004 eV$. As well the excess of $\gamma\gamma$ pair production  observed recently on LHC \cite{lhc} could be explained by the Higgs particle decay $h\rightarrow\gamma\gamma$ via virtual $\bar t t$ triangle loop if the Yukawa coupling is abnormally larger than the SM value, $\frac{g_t}{g_{t,SM}} = 1 \div 1.5$. All together it
entails the following bounds for the scales and couplings of our model,
\be
M > 3.5 TeV;\quad M_\ast > 3\cdot 10^8 GeV;\quad \kappa > 2\cdot 10^{- 15} .
\ee
Thus we conclude that the gravitational corrections on localization mechanism are indeed very small except for branon spectrum.  But the thickness of the brane may affect the high energy scattering processes already at the next LHC running and show up in appearance/disappearance processes, in particular in missing energy events \cite{rev12}, \cite{nesv} .
\section*{Acknowledgments}
We acknowledge the financial support by Grant RFBR 10-02-00881-a and by SPbSU grant 11.0.64.2010. One of us (A.A.) was partially supported by projects FPA2010-20807, 2009SGR502, CPAN (Consolider CSD2007-00042).

\end{document}